\documentstyle[aps,prb]{revtex}

\begin{document}

\title{The coil-globule transition of polymers of long rigid
monomers connected by flexible spacers}
\author{Richard P. Sear}
\address{FOM Institute for Atomic and Molecular Physics,
Kruislaan 407, 1098 SJ Amsterdam, The Netherlands\\
{\rm email: sear@amolf.nl}}

\maketitle

\begin{abstract}
A simple model of a polymer with long rigid segments which
interact via excluded volume repulsions and short ranged attractions is
proposed. The coil--globule transition of this model polymer is
strongly first order, the globule is crystalline and the
coil which coexists with the globule is swollen.
A virial expansion truncated at low order
is shown to provide a very poor approximation to the free energy and so
a cell theory is used to calculate the free energy of the globule.
\end{abstract}

%\newpage
\section{Introduction}

Typically, in a dilute polymer solution, each polymer chain
is spread over a very large volume.\cite{degennes}
The polymer chain forms a fractal
and if a box is drawn so as to enclose the chain, almost all of the volume
within the box is occupied by solvent, not by the monomers of the polymer.
This is the coil state of a polymer chain. The coil may be ideal,
in which case its radius of gyration scales with the number of
its monomers $N$ as $N^{1/2}$, or it may be swollen, in which case its
radius of gyration scales as $N^{3/5}$. The value $3/5$ of the
exponent is the Flory value,\cite{degennes} the true value is
slightly less.\cite{cloizeux90} 
However, there is a second state of the polymer: the globular
state, in which the radius of gyration scales as $N^{1/3}$.
As the radius of gyration scales with $N^{1/3}$, the volume
scales linearly with $N$, the density is then finite (except at
a second order transition to the coil state). The volume
occupied by a coil increases more rapidly than linearly with $N$ and
so as $N$ becomes very large the density tends to zero.
The state of a polymer
is in turn determined by the interactions between different
parts of the polymer. If they repel each other then the polymer is the
swollen coil state, if they attract each other strongly enough then the polymer
is in the globular state, whilst if the attractive and repulsive parts of the
interactions balance the polymer is ideal apart from
logarithmic corrections.\cite{degennes,cloizeux90}
By decreasing the temperature and so increasing the effect of the
attractive forces the polymer may be transformed from a coil to a globule.
A considerable amount of theoretical work has been devoted to attempting
to understand and describe the coil--globule transition.
However, this has concentrated almost exclusively on
coil--globule transitions driven
by the polymer's second virial coefficient
becoming negative.
\cite{lifshitz78,sanchez79,williams81,duplantier86,cloizeux90,grosberg92,ueda96,vasilevskaya97}
In the coil state, the interaction part
of the free energy is dominated by interactions between only two
parts of the polymer, which are accounted for by the polymer chain's
second virial coefficient (which is general different
from the second virial coefficient of the monomers).
So, as the second virial coefficient decreases to zero the interactions
effectively vanish and the coil becomes (almost) ideal.
Then on reducing the temperature further the second virial coefficient
becomes negative and the coil collapses into a globule.
The coil first becomes ideal before collapsing only when the collapse
is driven by the change of sign of the polymer's second virial coefficient.
Here, we show that when the monomers are long and rigid the
coil--globule transition is not driven by the second virial coefficient
becoming negative. So, the polymer is {\it never} ideal.
When the temperature is reduced
the polymer transforms via a strongly first order transition from a swollen
coil to a dense crystalline globule.

The use of a virial expansion\cite{hansen86} in order to describe the
contribution of monomer--monomer interactions to the free energy of the
polymer is reasonable for spherical monomers with not too short
ranged attractive interactions.
However, if the monomers are long and rigid an additional length
scale, the length of the monomer $L$, is present.
Unless the range of the attraction
is as long or longer than this length $L$ then the attractive forces
between the monomers are highly anisotropic.\cite{schoot93,sear97,schoot92}
Then, their effect on the third and higher virial coefficients
is larger than on the second virial coefficient.\cite{schoot92}
Long rigid monomers behave in
a similar way to hard spheres
with a short ranged attraction.
\cite{stell91,borstnik94,bolhuis94a,bolhuis94b,tejero95}.
Stell\cite{stell91,borstnik94} has shown that,
as the range of attraction between the spheres tends to zero,
the virial expansion becomes pathological
at temperatures such that the attractive interactions make a
nonnegligible contribution to the second virial coefficient.
Here, we show that the same occurs for long rods as the range of attraction
goes to zero.
Unless the renormalisation
required\cite{degennes} on going from the virial coefficients
of the monomers to the effective virial coefficients of a polymer
completely changes the behaviour of the virial series,
the coil--globule transition cannot be driven by the
renormalised second virial coefficient becoming negative.

The DNA double helix
is a rather rigid polymer, its persistence length\cite{vroege92}
is $\sim 50$nm\cite{merchant94} which is 25 times its diameter of
$\sim 2$nm. The interaction between DNA helices has a much shorter range than
50nm\cite{merchant94,israelachvili92,leikin91,odijk93}
and so we would expect the
behaviour of a DNA molecule of length $\gg 50$nm,
to be qualitatively similar to that of our model polymer chain.
This expectation is borne out,
the coil--globule transition of DNA, as for our model, is first order.
\cite{ueda96,vasilevskaya97,vasilevskaya95}

\section{The model polymer chain}

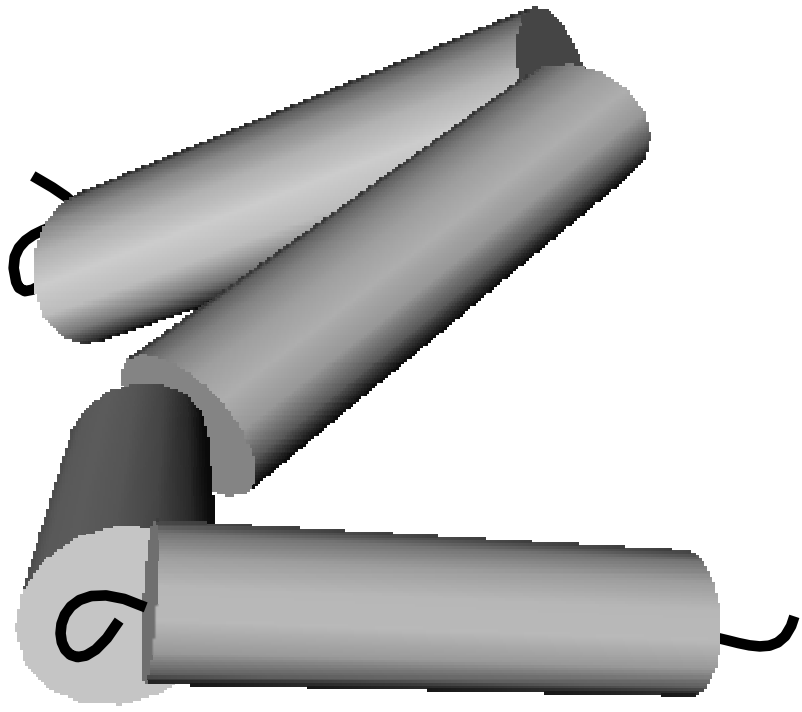
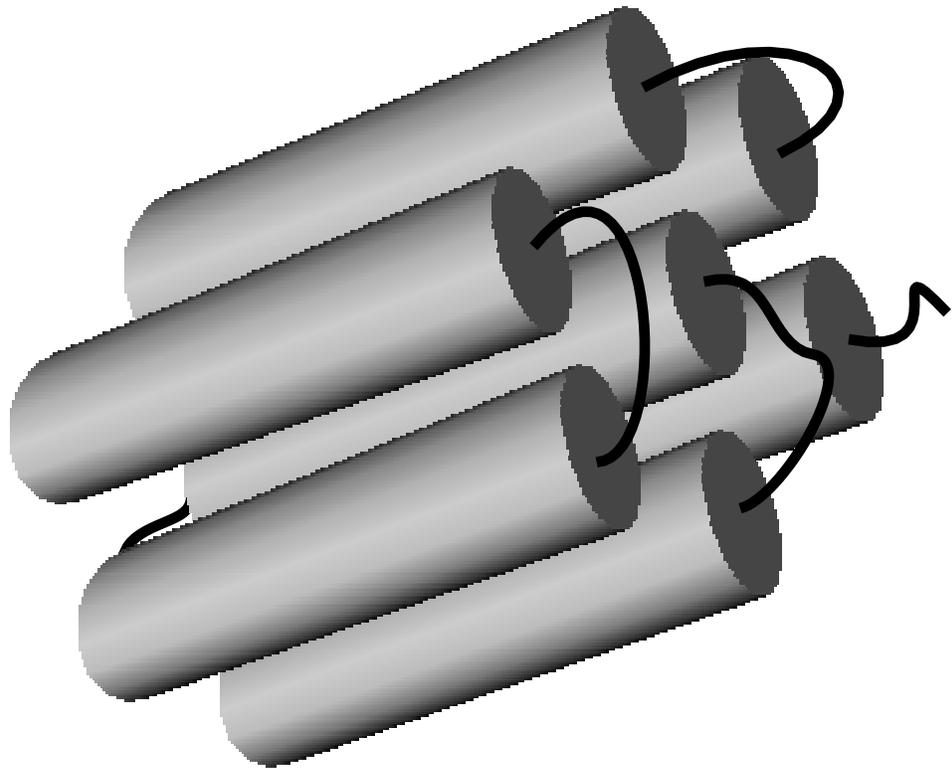
\begin{figure}
\caption{
Part of a polymer chain consisting
of $N$ rigid cylindrical monomers connected by flexible spacers;
a) shows the chain in the coil state and b) in the globule state.
The cylindrical monomers
are shown with a diameter smaller than $D$ in order to be able to
see the hexagonal packing in the globule state. The curved black
lines represent the spacers.
}
\label{fig1}
\end{figure}

The polymer is a linear chain of $N$ cylindrical monomers joined with
$(N-1)$ flexible spacers, see Fig. (\ref{fig1}).
A spacer is flexible enough to allow
the two cylinders which it joins to adopt any relative orientation with
equal probability and to lie side-by-side,
however, its maximum length is much less that the length $L$ 
of a cylindrical monomer.
We will always consider $N$ to be infinite and
so the coil--globule transition is a phase transition.
Real polymers are, of course, always finite but we do not consider the
effect of finite size here.

The cylindrical monomers are of length $L$, and have
a hard core of diameter $D$; when two cylinders overlap the energy
of interaction is positive and infinite.
The attractive part of the potential is sticky in the sense that its
range is close to zero, it is
only nonzero between a pair of cylinders if they almost touch for all of their
length, for this to be true they must be almost parallel and their centres of
mass must be almost side-by-side, i.e., only a little more than $D$ apart.
We will define what we mean by almost below when we define the potential.
When the attractive part of the potential is nonzero it equals $-\epsilon$.
The interaction $u$ between two cylinders,
1 and 2, is a function of the separation ${\bf r}_{12}$ of the
centres of mass of the cylinders, and of the angle
$\gamma_{12}$ between the centrelines of the two cylinders.
Then, $u({\bf r}_{12},\gamma_{12})$ is defined by
\begin{equation}
\begin{array}{ll}
\mbox{if}~~~~D<r_{12}<(1+\delta)D,~~~|z_{12}|<\delta D,~~~ \gamma_{12}<\delta
&~~~~~ u=-\epsilon\\
\mbox{else if the hard cores overlap}~~~~
&~~~~~ u=\infty\\
\mbox{otherwise}~~~~
&~~~~~ u=0,
\end{array}
\label{pot}
\end{equation}
where the vector ${\bf r}_{12}=(r_{12},z_{12},\phi_{12})$ in cylindrical
polar coordinates.
The origin of the coordinates is fixed on the
centre of mass of cylinder 1 and their $z$-axis is along its axis.
The parameter $\delta\ll 1$ defines how far apart the cylinders
can be and still interact via the attractive part of their interactions.
The limit $\delta\rightarrow0$
defines a zero range potential, a generalisation to a cylinder of the
sticky sphere potential of Baxter.\cite{stell91,baxter68}
The dependences of the attractive part of the potential $u$ on $\gamma_{12}$
and on the components of ${\bf r}_{12}$
are independent; this is rather artificial
but allows a simple evaluation of the second virial coefficient
and as $\delta$ is near zero the details of the attractive interaction
make little difference to the behaviour of the polymer.

In Eq. (\ref{pot}),
$\delta$ and $\epsilon$ are just adjustable parameters.
However, for an attractive interaction of range $\sim D$ between rods of
length $L$ the energy of interaction of two parallel rods scales linearly
with $L/D$.\cite{schoot92,israelachvili92} This produces a very deep well,
or a large $\epsilon$ in terms of our model. The energy continues to
scale linearly with $L/D$ as long as the angle between the two rods
is $\lesssim D/L$.
Thus, the $L/D\rightarrow\infty$ limit of the potential models
considered by van der Schoot and Odijk\cite{schoot92} is very similar
to the $\epsilon\rightarrow\infty$,
$\delta\rightarrow0$ and $\epsilon\delta\rightarrow$constant
limit of Eq. (\ref{pot}).

\section{The coil state}

The density of a coil tends to zero as $N$ tends to infinity.
Thus, the only interactions between distant parts of the chain that count
are pair interactions; the density is so low that the probability of three
different parts of the chain coming close enough to interact is
vanishingly small.\cite{degennes,cloizeux90}
By distant parts of the chain we mean monomers which are not close together
along the chain, i.e., are separated along the chain by many other monomers.
As only pair interactions are important the excess free energy
depends on only one parameter,
an effective second virial coefficient $B_2^r$.
This is not the same as the second virial coefficient between
monomers, $B_2$. The effective coefficient $B_2^r$ is obtained by
renormalising $B_2$; this `mixes in' some of the higher virial coefficients
in with $B_2$ to produce $B_2^r$. So, we start by determining $B_2$
and then go on to estimate the higher virial coefficients.

In the limit of the temperature tending to infinity,
the polymer chain is just a chain
of hard cylinders and flexible spacers.\cite{khokhlov85}
Then, $B_2=(\pi/4)L^2D$.\cite{vroege92,onsager49}
The higher virial coefficients are all very small if $L/D\gg1$.
\cite{vroege92,onsager49} Thus interactions between pairs of the monomers
are close to being independent, and so if two distant (along the chain)
monomers are interacting the probability of any of the monomers
which are adjacent to either of these monomers
interacting with either of the original pair of monomers is very low.
As it is these interactions which renormalise $B_2$, we conclude
that the renormalisation required to derive $B_2^r$ from $B_2$ is weak and
so $B_2^r\simeq B_2$.
The pair interactions cause the coil
to swell so that its radius of gyration scales as $N^{3/5}$, not as
$N^{1/2}$ as an ideal coil does.
However, they contribute an amount much less than $T$ to the free energy
per monomer.
$T$ is the temperature in energy units.
This can be seen if we examine the interactions between
distant and between adjacent monomers. The interaction between distant
pairs of monomers, i.e., pairs of monomers which are not neighbours
or next-nearest neighbours etc. along the chain, is characterised
by the Fixman parameter\cite{degennes,cloizeux90,doi86} $z\sim B_2/L^3N^{1/2}$.
If we divide this by $N$ to obtain the contribution per monomer we see that
the contribution tends to zero as $N$ tends to infinity.
As for the interaction between adjacent monomers;
it is easy to see that a monomer only
very weakly restricts the space available to an adjacent monomer. Thus, the
contribution to the free energy from interactions between monomers is small.
Then the free energy of our chain is to a good approximation the free
energy of an ideal chain. If the flexible spacers allow each monomer
to move freely in a phase space of volume $D^3$ then the free energy
per monomer of the coil $a$ is
\begin{equation}
a(T)=-T\ln D^3.
\label{ac}
\end{equation}
Of course there is also a term from the momenta but we neglect this as
it is the same in all phases.
Fixing the volume of phase space to be $D^3$ is consistent with short
tethers of length not much larger than $D$.

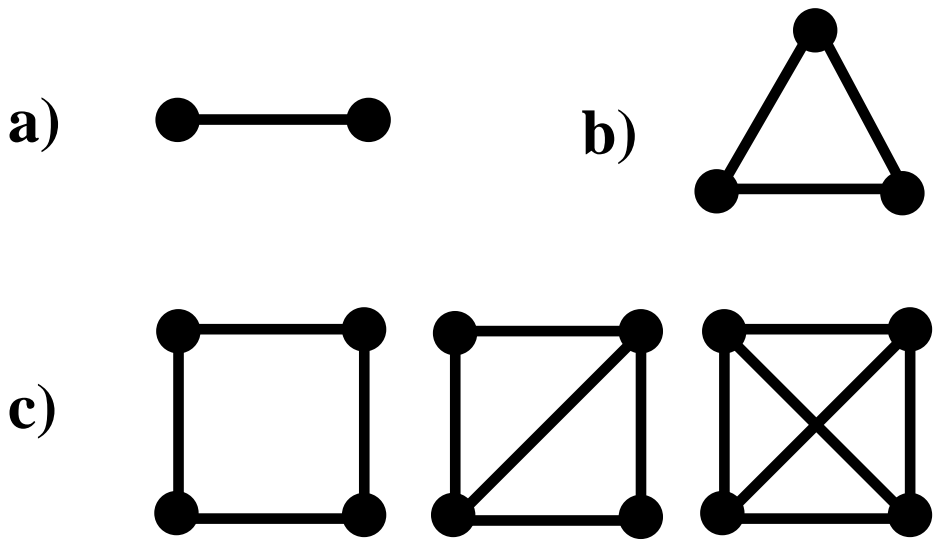
\begin{figure}
\caption{
The diagrams for the second, a), third, b), and
fourth, c), virial coefficients. The black circles represent the rods
and the lines represent Mayer f functions, which are $\exp[-u/T]-1$.
A diagram with $n$ circles is an integral over the coordinates
of $n$ rods, interacting by the Mayer f functions shown in the diagram,
divided by a symmetry number.
}
\label{fig2}
\end{figure}

In order to consider the effect of attractive interactions we start by
studying the second and third virial coefficients of sticky cylinders.
In fact the contribution of pairwise interactions to the
free energy of a coil involves not $B_2$ but
a renormalised second virial coefficient $B_2^r$.\cite{degennes}
The difference between the two
will be considered after the unrenormalised coefficients have been calculated.
The second virial coefficient is\cite{hansen86,vroege92}
\begin{equation}
B_2(T)=-\frac{1}{32\pi^2}\int
\left(\exp[-u({\bf r}_{12},\gamma_{12})/T]-1\right)
d{\bf r}_{12}d\Omega_1d\Omega_2,
\label{b2def}
\end{equation}
where $d\Omega_i=\sin\theta_id\theta_id\phi_i$.
The diagrammatic representation of $B_2$ is given by Fig. (\ref{fig2}a).
See Ref. \onlinecite{hansen86} for an introduction to the theory of diagrams.
The integral in Eq. (\ref{b2def}) is
straightforward. For $\gamma>\delta$ the interaction is just that between
two hard rods, whose $B_2$ is well known.\cite{vroege92,onsager49}
The contribution of the attractive part of the interaction is evaluated
by aligning cylinder 1 with the $z$-axis,
then $\gamma_{12}=\theta_2$,
and as the integration is restricted to small angles
$\sin\theta_2\simeq\theta_2$. So,
\begin{equation}
B_2(T)=\frac{\pi}{4}L^2D-\frac{1}{4}
\exp[\epsilon/T]
\int_D^{D+\delta D} 2\pi r_{12}dr_{12}\int_{-\delta D}^{\delta D}dz_{12}
\int_0^{\delta}\theta_2 d\theta_2,
\end{equation}
where cylindrical polar coordinates have been used for ${\bf r}_{12}$.
We have assumed that $L/D\gg1$ and $\epsilon/T\gg1$.
\begin{equation}
B_2(T)=\frac{\pi}{4}L^2D-
\frac{\pi}{2} D^3\delta^4\exp[\epsilon/T].
\label{b2}
\end{equation}
This expression for $B_2$ may be compared with those derived by
van der Schoot and Odijk\cite{schoot92} for more realistic potentials.
If in Eq. (\ref{b2}), $\delta$ is replaced by $D/L$ and $\epsilon$
multiplied by $L/D$, then it becomes essentially the same as
Eq. (3.7) of Ref. \onlinecite{schoot92}.

The twelfth virial coefficient of sticky spheres diverges to minus
infinity at all temperatures at which the attractive interactions make a
nonnegligible contribution to the second virial coefficient.\cite{stell91}
Here, we show that the third virial coefficient $B_3$ of sticky cylinders
diverges to minus infinity at all temperatures for which the attractive
interactions make a nonnegligible contribution to $B_2$, in the limit
that $\delta$ tends to zero.
$B_3$ is defined by\cite{hansen86}
\begin{equation}
B_3(T)=-\frac{1}{384\pi^3}\int\prod_{ij=12,13,23}
\left(\exp[-u({\bf r}_{ij},\gamma_{ij})/T]-1\right)
d{\bf r}_{12}d{\bf r}_{13}d\Omega_1d\Omega_2d\Omega_3 ,
\end{equation}
where ${\bf r}_{13}$ is the vector between the centres of mass of rods 1
and 3, and where $\gamma_{13}$ and $\gamma_{23}$ are the angles
between rods 1 and 3, and 2 and 3, respectively.
The diagrammatic representation of $B_3$ is given by Fig. (\ref{fig2}b).
Only the scaling of $B_3$ with $\epsilon$ and $\delta$
is required, it is
\begin{equation}
B_3(T)\sim L^3D^3 - D^6\delta^9\exp[3\epsilon/T].
\label{b3}
\end{equation}
The first term is from the hard core.\cite{vroege92,onsager49}
There are three interactions between the three rods, this gives
the factor of 3 in the exponential of the second term of Eq. (\ref{b3}).
The integration over the interaction of the first rod with each of the
the second and third rods gives a factor of $\delta^4$, then
as the rods are already restricted to be nearly parallel and
the second and third rods are already restricted to being close to the first
rod the integration over the interaction between the second and third rods
gives only a factor of $\delta$.
In the limit of $\delta\rightarrow0$ the second term of
$B_2$, Eq. (\ref{b2}),
is finite if $\epsilon$ diverges as
$\epsilon\rightarrow\ln\delta^{-4}$.
Then the attractive interactions contribute a finite amount to the value
of $B_2$.
Putting this dependence
of $\epsilon$ into Eq. (\ref{b3}) results in the attractive
term in $B_3$ tending to minus infinity
as $\delta\rightarrow0$.
The divergence of the twelfth virial coefficient of sticky hard spheres
is suppressed by the presence of even very weak
polydispersity in the diameter of the spheres. In contrast,
polydispersity in the diameter $D$ of sticky rods
has little effect: $B_3$ still diverges.

Although we cannot consider all virial coefficients we are able to
find a trend in their behaviour. 
The fourth virial coefficient $B_4$ is the sum of
the 3 diagrams of Fig. (\ref{fig2}c).\cite{hansen86}
The most highly connected of these
(with 6 bonds) does not correspond to a realisable cluster;
\cite{stell91,wertheim84}
it is not possible for all of the 4 rods to interact simultaneously
with all of the
other 3 rods via the attractive part of the interaction. Therefore,
it is the diagram with 5 bonds (the middle one of Fig. (\ref{fig2}c))
which dominates at low temperature.
Five bonds means a factor of $\exp[5\epsilon/T]$.
The region of integration over which all 5 bonds are simultaneously
present scales as $\delta^{14}$.
This result may be derived if we start from the $B_3$ integral
and add an extra rod, which interacts
with 2 of the existing rods.
The integral over the 3 rods in $B_3$ yields a factor of $\delta^9$,
then restricting the
orientation of the fourth rod introduces an additional factor of $\delta^2$.
The centre of mass of the fourth rod must be restricted in all 3 directions
in order that it interact with 2 other rods;
this gives a factor of $\delta^3$.
So, we can find the
order of the diagrams in the $n$th virial coefficient with respect
to $\exp[\epsilon/T]$ and $\delta$ by starting from diagrams in the $(n-1)$th
coefficient and adding another rod such that the cluster is still
realisable, i.e., that all the rods can simultaneously interact
via all the bonds in the diagram. However, this quickly becomes tedious
as $n$ increases. We have found that the part of $B_4$ due
to the attractive interactions scales as $\exp[5\epsilon/T]\delta^{14}$,
and so becomes non-zero at temperatures such that both $B_2$ and $B_3$
are still at their $T=\infty$ limits. The trend seems clear, 
$B_n$ starts to differ from its $T=\infty$ limit at a temperature
which increases with $n$.

So far, we have just studied the virial coefficients of
sticky cylinders. As the third
virial coefficient diverges at all temperatures
such that the attractive interactions influence $B_2$,
a fluid of sticky cylinders
is unstable at these temperatures, just as sticky spheres are.\cite{stell91}
However, the main object of our study is a polymer chain
consisting of an sequence of $N$ of these sticky cylinders.
In a polymer coil the effective second virial coefficient is
not that of the monomers $B_2$ but is $B_2$ renormalised
by the higher coefficients, $B_2^r$.\cite{degennes}
The temperature at which the coil is ideal is the temperature at which
$B_2^r$ is zero.
At low temperatures the higher virial coefficients are much larger
than $B_2$ and so it is reasonable to expect the renormalisation
to produce a renormalised $B_2^r$ which differs substantially from the
$B_2$ of Eq. (\ref{b2}). Indeed, there is no reason why
the scaling of $B_2^r$ and $B_2$ with $\delta$ and $\epsilon$ should be the
same. Unfortunately we do not know how to carry out this renormalisation.
In conclusion, at high temperatures we can approximate $B_2^r$ by $B_2$
and in the coil state $B_2^r$ is all we require.
The assumption that $B_2^r\sim B_2$ near the $T=\infty$ limit is
reasonable because then
for $L/D\gg1$ the higher virial coefficients are all very small. At low
temperatures we have little idea of the size of $B_2^r$ and so
we are unable to construct a virial expansion of the free energy
of a coil. The crossover temperature between these two regimes will
be estimated in Section V.

\section{The globular state}

A globule, because it is denser than a coil, has a lower entropy. This is
compensated by its lower energy due to attractive interactions between
the monomers of the polymer. Our rodlike monomers only attract each other
when they are side-by-side and nearly parallel. Given this, it is easy
to see that a monomer has the lowest energy when it is
closely surrounded by 6 other monomers, all of which are nearly parallel
and have their centres of mass almost in a plane. Then the monomer
interacts strongly with 6 other monomers, the maximum number possible.
The lowest energy state of the polymer is then just a state which allows
every monomer to be surrounded in this way by 6 others. This occurs
in a solid formed from layers of cylinders arranged in 2-dimensional
hexagonal lattices.
It is not surprising that the lowest energy is achieved in a solid but
it should be noted that our monomers have to lie side-by-side to attract
each other and so the energy of any isotropic phase will be much higher
than that of a phase with orientational ordering. In addition the
orientational ordering must be very pronounced, the angle the monomers
make with the director should not be more than $\sim\delta$.
The restriction that the centres of mass must be within $(1+\delta)D$
in order for them to attract each other forces the density to be very high,
roughly a fraction $1/(1+\delta)^2$ of the density of close
packing --- the maximum possible density.
A globule with a volume fraction of, say, 1\% would necessarily have
a much higher energy due to the very short range of the attractions.

As we expect a globule of such high density,
a cell theory \cite{tejero95,hill60}
is used to estimate its free energy.
We consider a very dense layer of cylinders with their
centres of mass on a 2-dimensional triangular lattice; the
lattice constant is $(1+\delta/2)D$.
As the interlayer interactions are very weak in our model we do not
consider the interactions between monomers in different layers.
Fig. (\ref{fig1}b) is a schematic of part of such a layer. 
In a cell theory calculation, the
nearest neighbours of a particle are fixed at their places in a perfect
lattice and the partition function $q_1$ of a particle trapped in the
cell formed by them evaluated.
So, the centres of mass of the
six neighbours of the particle are fixed at positions $(1+\delta/2)D$
away from the centre
of the cell of the particle and in the
$xy$-plane of our coordinate system and their centrelines are parallel
to its $z$-axis.
The cylinder in the cell has 5 degrees of freedom,
3 translational and 2 rotational.
As our model is defined so that the orientational and translational
degrees of freedom are decoupled, estimating the volume of phase
space available to the rod is straightforward.
The orientational degrees of freedom give a factor of
$(1/2)\delta^2$, as for $B_2$.  We
approximate the volume available to the centre of mass of the
rod by a hexagonal prism with its axis along the $z$ axis.
The height of the prism is $2\delta D$ and the distance from its
centre to any of its points is $(1/2)\delta D$; its volume is then
$0.75\sqrt{3}(\delta D)^3$. So,
\begin{equation}
q_1\simeq\frac{3\sqrt{3}}{8} D^3\delta^5 \exp[3\epsilon/T]
\label{q1}
\end{equation}
and the free energy per monomer $a$ is
\begin{equation}
a(T)=-T\ln q_1\simeq -3\epsilon-
T\ln\left[\frac{3\sqrt{3}}{8}D^3\delta^5\right].
\label{ah}
\end{equation}
The integral over the momenta coordinates
has been neglected; it is
the same in all phases and so has no effect on the phase behaviour.
In taking $a(T)$ to be the free energy per monomer of our polymer we have
neglected the entropy associated with the number of ways in which the
monomers can be connected by the spacers; this is of order unity.

\section{The coil--globule phase transition}

In Section II, we showed only that the
when $B_2$ becomes negative the higher coefficients are
already negative, and infinite.
Now that we have an estimate of the free energy of the globule we can
determine the stability of the coil with respect to the globule when
$B_2$ becomes negative.
In the limit of $\delta\rightarrow\infty$ the second term of
$B_2$, Eq. (\ref{b2}),
is finite if $\epsilon$ diverges as
$\epsilon\rightarrow\ln\delta^{-4}$. Putting this dependence
of $\epsilon$ into Eq. (\ref{ah}) for the free energy of a globule,
results in the free energy of the globule
tending to $-\infty$. Thus, for cylinders
with a very short ranged, sticky, attraction the coil is unstable
with respect to the globule when the (unrenormalised) interaction
between a pair of monomers becomes attractive.
The coil is only stable at temperatures such that $B_2$ equals its
$T=\infty$ limit.

Now, as the free energy per particle of the coil state is constant
(at high temperature)
and given by Eq. (\ref{ac}),
we can estimate
the temperature of the coil--globule transition $T_{cg}$ by
equating the free energy of the globule state, Eq. (\ref{ah}),
to that in the coil state, Eq. (\ref{ac}),
\begin{equation}
\frac{T_{cg}}{\epsilon}\simeq\frac{-3}
{\ln\left[\frac{3\sqrt{3}}{8}\delta^5\right]}.
\label{tcg}
\end{equation}
Above $T_{cg}$ the coil is more stable than the globule, below $T_{cg}$ the
reverse is true. At $T_{cg}$, the coil collapses via a strongly first order
transition to form a dense globule.
This is qualitatively different from the situation
when the polymer is flexible, there the
radius of gyration changes continuously as the temperature is varied.
\cite{lifshitz78,williams81,grosberg92}
{\it A posteriori} justification for the correctness of Eq. (\ref{tcg})
is provided by the fact that it predicts a $T_{cg}$ at which
both $B_2$, Eq.(\ref{b2}), and $B_3$, Eq. (\ref{b3}), are equal to their
$T=\infty$ limits.
So, unless $B_2^r$ at $T_{cg}$ contains large contributions from higher order
virial coefficients the coil is close to its $T=\infty$ limit at $T_{cg}$
and so the free energy per monomer is close
to the value of Eq. (\ref{ac}).
Physically, large contributions from high order virial coefficients
corresponds to clustering, i.e., significant numbers of monomers in the
coil would be part of clusters of 4 or more monomers.\cite{sear97}
At equilibrium, there seems no reason for small clusters to be favoured,
if the attractive interactions are strong enough to bind a few
monomers into a cluster then they are strong enough to form a
macroscopic cluster --- the globular phase.

The temperature $T_{cg}$ is an approximation to the temperature
at which a fluid of the (unpolymerised) monomers becomes unstable with respect
to a dense solid.\cite{schoot93,sear97}
It is an overestimate as free monomers possess
translational entropy in the fluid phase which stabilises the fluid.
On the basis of our analysis here
we expect a fluid of long rodlike objects with short ranged attractions
to undergo a strongly first order transition to a very dense phase, 
when the attractions are made sufficiently strong.
This appears to happen for short, $\sim 50$nm, lengths of DNA.
\cite{wissenburg94}
Wissenburg et al.\cite{wissenburg94} have observed that, on increasing
the concentration of the DNA, it suddenly forms dense aggregates.

\section{Conclusion}

The polymer never forms an ideal coil, at temperatures above
$T_{cg}$ it exists as a swollen coil and below $T_{cg}$ it forms a
crystalline globule. The coil--globule transition is strongly first order.
A cell theory was used to calculate the free energy of the globular state
and so to find
the coil--globule transition temperature $T_{cg}$;
the virial expansion for the free energy of our
polymer having been shown to be useless at low temperatures.
The coil--globule transition of a flexible polymer is second order,
\cite{duplantier86,grosberg92,grassberger95,wittkop96} the radius of gyration
varies continuously with temperature. This globule is liquid-like,
there is no orientational or translational ordering.
Clearly, as the stiffness of the
monomers is increased there is a change in the character of the coil--globule
transition; this change has been observed in simulations of lattice models.
\cite{kolinski86,doniach96,bastolla,doye}
Experimentally, data is only available for highly flexible polymers, such
as polystyrene\cite{williams81,grosberg92,sun80}
and DNA.\cite{ueda96,vasilevskaya97}, which is semiflexible
The data for flexible chains seems consistent with theoretical
results based on virial expansions
\cite{lifshitz78,williams81,duplantier86,grosberg92} and with
simulations of flexible chains on a lattice.\cite{grassberger95,wittkop96}
However, DNA forms dense \cite{ueda96,vasilevskaya97,lerman71,ubbink95}
globules in which the DNA helices
are hexagonally ordered.
We have studied, using theory, a molecular model
which also forms dense ordered globules.
The major difference between our model and
DNA is that DNA is semiflexible, which means that the helix bends continuously,
whereas our chain consists of completely rigid monomers, which
cannot bend at all, and
spacers, at which the chain can bend freely.
We hope to be able to incorporate the effect
of continuous flexibility in future work.

It is a pleasure to thank J. Doye
for stimulating discussions and a careful reading of the manuscript.
I would like to thank The Royal Society for the award of a fellowship
and the FOM institute AMOLF for its hospitality.
The work of the FOM Institute is part of the research program of FOM
and is made possible by financial support from the
Netherlands Organisation for Scientific Research (NWO).

%\newpage

\end{document}